\documentclass[preprintnumbers, amsmath, amssymb, showpacs,twocolumn, superscriptaddress, nofootinbib]{revtex4-1}

\usepackage{graphicx}
\usepackage{epsfig}
\usepackage{dcolumn}
\usepackage{bm}
\usepackage{amsfonts}
\usepackage{subfigure}
\usepackage{color}

\newcommand{\be}{\begin{equation}}
\newcommand{\ee}{\end{equation}}

\newcommand{\bea}{\begin{eqnarray}}
\newcommand{\eea}{\end{eqnarray}}

\def\gsim{ \lower .75ex \hbox{$\sim$} \llap{\raise .27ex \hbox{$>$}} }
\def\lsim{ \lower .75ex \hbox{$\sim$} \llap{\raise .27ex \hbox{$<$}} }

\begin{document}


\title{$f(T)$ gravity after GW170817 and GRB170817A}

\author{Yi-Fu Cai}
\email{yifucai@ustc.edu.cn}
\affiliation{CAS Key Laboratory for Research in Galaxies and Cosmology, Department of Astronomy, University of Science and Technology of China, Hefei 230026, China}
\affiliation{School of Astronomy and Space Science, University of Science and Technology of China, Hefei 230026, China}

\author{Chunlong Li}
\email{chunlong@mail.ustc.edu.cn}
\affiliation{CAS Key Laboratory for Research in Galaxies and Cosmology, Department of Astronomy, University of Science and Technology of China, Hefei 230026, China}
\affiliation{School of Astronomy and Space Science, University of Science and Technology of China, Hefei 230026, China}

\author{Emmanuel N. Saridakis}
\email{Emmanuel\_Saridakis@baylor.edu}
\affiliation{Department of Physics, National Technical University of Athens, Zografou Campus GR 157 73, Athens, Greece}
\affiliation{CASPER, Physics Department, Baylor University, Waco, TX 76798-7310, USA}

\author{LingQin Xue}
\email{xuelq@mail.ustc.edu.cn}
\affiliation{CAS Key Laboratory for Research in Galaxies and Cosmology, Department of Astronomy, University of Science and Technology of China, Hefei 230026, China}
\affiliation{School of Astronomy and Space Science, University of Science and Technology of China, Hefei 230026, China}

\begin{abstract}

The combined observation of GW170817 and its electromagnetic counterpart GRB170817A
reveals that gravitational waves propagate at the speed of light in high precision. We
apply the effective field theory approach to investigate the experimental consequences for
the theory of $f(T)$ gravity. We find that the speed of gravitational waves within $f(T)$
gravity is exactly equal to the light speed, and hence the constraints from GW170817 and
GRB170817A are trivially satisfied. The results are verified through the standard analysis
of cosmological perturbations. Nevertheless, by examining the dispersion relation and the
frequency of cosmological gravitational waves, we observe a deviation from the results of
General Relativity, quantified by a new parameter. Although its value is relatively small
in viable $f(T)$ models, its possible future measurement in advancing
gravitational-wave astronomy would be the smoking
gun of testing this type of modified gravity.
\end{abstract}


\maketitle

\section{Introduction}

Modified gravity \cite{Nojiri:2006ri,Capozziello:2011et} is a crucial direction that one
can follow in order to explain the early- and late-time phases of accelerated expansion
of the Universe, instead of the introduction of dark energy \cite{Copeland:2006wr,
Cai:2009zp} and the inflaton \cite{Guth:1980zm}. Most theories that are modifications of
General Relativity (GR) are based on various extensions of the standard Einstein-Hilbert
action, and thus lie within the curvature-based gravitational formulation, namely
$f(R)$ gravity \cite{DeFelice:2010aj}, $f(G)$ gravity \cite{Nojiri:2005jg}, etc.
Alternatively, one can construct gravity theories starting from the torsion based
formulation, and in particular from the Teleparallel Equivalent of General Relativity
(TEGR) \cite{ein28,ein28b, Weitzen28, Hayashi79, Pereira:book, Maluf:2013gaa}. In this
formulation the Lagrangian is the torsion scalar $T$, which is obtained by the contraction
of the torsion tensor, and thus one can use it to construct torsional extended theories,
namely $f(T)$ gravity \cite{Bengochea:2008gz, Linder:2010py}, $f(T_G)$ gravity
\cite{Kofinas:2014owa, Kofinas:2014daa}, etc (see \cite{Cai:2015emx} for
a comprehensive review). Although TEGR and GR are equivalent at the level of equations,
their modifications correspond to different theoretical developments, and therefore in
recent years theories of torsional modified gravity have attracted the interest of
physicists in the literature \cite{Ferraro:2006jd, Chen:2010va, Dent:2011zz, Wu:2010mn,
Cai:2011tc, Li:2011rn, Bamba:2012vg, Wu:2011kh, Ong:2013qja, Otalora:2013tba,
Bahamonde:2015zma, Rezazadeh:2015dza, Farrugia:2016qqe, Awad:2017tyz}.

To examine which modified gravitational theories amongst the huge zoo of proposals are
good candidates for the description of Nature, we resort to comparison with observations.
Apart from the standard observational data that one can use, including Supernovae (SN),
Cosmic Microwave Background (CMB), Baryonic Acoustic Oscillations (BAO), the growth rate,
the Hubble function, etc, the recent detection of gravitational waves (GWs) has opened a
new window in exploring the Universe. Furthermore, the simultaneous observation of the
associated electromagnetic counterparts in case of binary neutron star mergers, such as
GW170817 detected by the LIGO-VIRGO collaboration \cite{TheLIGOScientific:2017qsa} and
GRB170817A by the Fermi Gamma-ray Burst Monitor \cite{Goldstein:2017mmi}, imposes
strong constraints on the speed of GWs, namely $|c_g/c-1|\leq 4.5\times 10^{-16}$
\cite{Monitor:2017mdv}. This bound is crucial in constraining or excluding large classes
of modified gravity theories, since in gravitational modifications the GWs propagate with
a speed that is in general different than the light speed \cite{Ezquiaga:2017ekz,
Baker:2017hug}.

In the present work we are interested in calculating the speed of GWs in $f(T)$ gravity.
We will apply the effective field theory (EFT) approach which allows to analyze the
perturbations in a systematic way and separately from the background evolution, and we
will
verify the results through the standard scalar, vector and tensor perturbations. As we
shall see, we find that in $f(T)$ gravity the GWs propagate at the speed of light, and
thus the GW170817 bounds are trivially satisfied. The structure of the manuscript is as
follows. In Section \ref{fTgravity} we briefly review $f(T)$ gravity at the background
level, and in Section \ref{SecionGravwaves} we perform the standard perturbation analysis
around a cosmological background. In Section \ref{EFTanalysis} we apply the EFT approach
to $f(T)$ gravity and we extract the propagation equation and speed of GWs. Finally,
Section \ref{Seciontconclusions} is devoted to conclusions and discussion.

\section{$f(T)$ gravity}
\label{fTgravity}

Let us briefly review the $f(T)$ gravitational theory  \cite{Cai:2015emx}.  In torsional
and teleparallel gravity one uses the tetrad fields $e_A^\mu$ as the dynamical variables,
which are defined at each point of the manifold as a base of orthonormal vectors
$e^{\mu}_A$, where $A,B,C...=0,1,2,3$ label the tangent spacetime coordinates, while $\mu,
\nu, \rho ... = 0, 1, 2, 3$ are the spacetime coordinates. Furthermore, a co-tetrad
$e_{\mu}^A$ is defined through $ e^{\mu}_A e^A_{\nu} = \delta^{\mu}_{\nu}$ and $e^{\mu}_A
e^B_{\mu} = \delta^A_B$.

In order to describe the orthogonality and normalization of tetrad fields one introduces
the tetrad metric $\eta_{AB} = \eta^{AB} = diag(-1, 1, 1, 1)$ \footnote{Note that in
order to be closer to standard gravity in this work we construct torsional gravity
following the mostly-plus signature as in \cite{Kofinas:2014daa}, instead of following
the mostly-minus signature as in \cite{Bengochea:2008gz, Linder:2010py,Chen:2010va},
which leads to some sign differences in the intermediate expressions. Definitely, the
cosmological equations and observables are the same in both conventions.}, and thus the
space-time metric can be reconstructed as
\begin{align}
\label{metric}
 g_{\mu\nu}=\eta_{AB}e^A_{\mu}e^B_{\nu} ~.
\end{align}
In teleparallel gravity, one uses the Weitzenb\"{o}ck connection
$\hat{\Gamma}^{\lambda}_{\ \mu\nu}
\equiv e^{\lambda}_A \partial_{\nu} e^A_{\mu} = - e^A_{\mu} \partial_{\nu}
e^{\lambda}_{A}$, which
is a connection leading to zero curvature but non-zero torsion. The resulting torsion
tensor is
\begin{align}
\label{torsiontensor}
 T^{\lambda}_{\ \mu\nu} \equiv \hat{\Gamma}^{\lambda}_{\ \nu\mu}
-\hat{\Gamma}^{\lambda}_{\ \mu\nu}
= e^{\lambda}_A(\partial_{\mu} e^A_{\nu} - \partial_{\nu} e^A_{\mu}) ~,
\end{align}
and therefore one can construct the torsion scalar through its contractions, namely
\begin{equation}
\label{torsiscal}
 T\equiv\frac{1}{4} T^{\rho \mu \nu} T_{\rho \mu \nu} + \frac{1}{2}T^{\rho \mu \nu }T_{\nu
\mu\rho}
- T_{\rho \mu }^{\ \ \rho }T_{\ \ \ \nu }^{\nu \mu} ~.
\end{equation}

As is well known, the Levi-Civita connection $\Gamma^{\sigma}_{\ \mu\nu}$ is related to
any other
connection, and thus to Weitzenb\"{o}ck connection
too, through
\begin{equation}
\label{LR}
 \hat{\Gamma}^{\rho}_{\ \mu\nu} = \Gamma^{\rho}_{\ \mu\nu} + {\cal{K}}^{\rho}_{\ \mu\nu}
~,
\end{equation}
where the  contorsion tensor writes as
\begin{align}
\label{KK}
 {\cal{K}}^{\rho}_{\ \mu\nu}\equiv\frac{1}{2}(T_{\mu\ \nu}^{\ \rho} + T_{\nu\ \mu}^{\
\rho} - T^{\rho}_{\ \mu\nu}) ~.
\end{align}
Similarly, the covariant derivative of a quantity $A_{\mu}$ with respect to the
Levi-Civita
connection $\nabla_{\nu}$ is related to its covariant derivative with respect to the
Weitzenb\"{o}ck connection $ \hat{\nabla}_{\nu}$ through
\begin{equation}
\label{DR}
 \hat{\nabla}_{\nu}A_{\mu}=\nabla_{\nu}A_{\mu}-{\cal{K}}^{\rho}_{\mu\nu}A_{\rho} ~.
\end{equation}
Hence, since the curvature (Riemann) tensor corresponding to the Levi-Civita connection is
\begin{align}
 R^{\rho}_{\ \lambda\mu\nu} = \partial_{\mu} \Gamma^{\rho}_{\ \lambda\nu} +
\Gamma^{\rho}_{\ \sigma\mu} \Gamma^{\sigma}_{\ \lambda\nu}
 - \partial_{\nu} \Gamma^{\rho}_{\ \lambda\mu} - \Gamma^{\rho}_{\ \sigma\nu}
\Gamma^{\sigma}_{\ \lambda\mu} ~,
\end{align}
one can straightforward derive the relation
\begin{align}
\label{RT}
 R=-T+2\nabla_{\mu}T^{\mu} ~.
\end{align}
Here, $R$ is the Ricci scalar corresponding to the Levi-Civita connection, $T$ is the
torsion scalar (\ref{torsiscal}) corresponding to the Weitzenb\"{o}ck connection, and
$T^{\mu}$ is the contraction of the torsion tensor, defined as
$T_{\mu}\equiv T^{\nu}{}_{\nu\mu}$.

In teleparallel gravity one uses the above torsion scalar as the Lagrangian of the theory,
in a similar way to the use of the Ricci scalar as the Lagrangian of general relativity.
Due to relation (\ref{RT}) one can immediately see that the two theories will
be completely equivalent at the level of equations, and that is why this theory
is called TEGR. Nevertheless, one can use TEGR as a base of extended
gravity. Inspired by the $f(R)$ extensions of GR, one can generalize $T$ to
a function $f(T)$, resulting to $f(T)$ gravity,
which is characterized by the action
\begin{align}
\label{fT}
 S=\int d^4x e\, \frac{M_P^2}{2}f(T) ~,
\end{align}
where $e = \text{det}(e_{\mu}^A) = \sqrt{-g}$ and with $M_P$ the Planck mass in units
where the
light speed is set to $c=1$. Varying the above action with
respect to the tetrads we extract the field equations as
\begin{align}
\label{fieldeqs}
 & e^{-1}\partial_{\mu} (ee_A^{\rho}S_{\rho}{}^{\mu\nu}) f_{T} + e_A^{\rho}
S_{\rho}{}^{\mu\nu} \partial_{\mu}({T}) f_{TT} \nonumber\\
 & - f_{T} e_{A}^{\lambda} T^{\rho}{}_{\mu\lambda} S_{\rho}{}^{\nu\mu} + \frac{1}{4} e_
{A}^{\nu} f(
{T}) 
 = 4 \pi G e_{A}^{\rho} \Theta_{\rho}{}^{\nu} ~,
\end{align}
where $f_{T}=\partial f/\partial T$, $f_{TT}=\partial^{2} f/\partial T^{2}$, and with
$\Theta_{\rho}
{}^{\nu}$  denoting the matter energy-momentum tensor. For convenience, in the above
equation we have introduced the
``super-potential''
\begin{eqnarray}
\label{superpot}
 S_\rho^{\:\:\:\mu\nu} \equiv \frac{1}{2} \Big( {\cal{K}}^{\mu\nu}_{\:\:\:\:\rho} +
\delta^\mu_\rho \: T^{\alpha\nu}_{\:\:\:\:\alpha} - \delta^\nu_\rho \:
T^{\alpha\mu}_{\:\:\:\:\alpha} \Big) ~.
\end{eqnarray}

\section{GWs in $f(T)$ gravity}
\label{SecionGravwaves}

In this section we investigate cosmological GWs generated in $f(T)$ gravity. Since the
dynamical variables are the four vector tetrad fields, instead of the symmetric metric
field,  we need to consider all the $16$ components of the tetrads instead of the $10$
components of the metric tensor. Definitely, comparing with the metric tensor, which has
only coordinate indices, we should note that the tetrad $e^A_{\mu}$ has additional tangent
space-time indices, and therefore the local Lorentz invariance will release 6 extra
degrees of freedom \cite{Izumi:2012qj,Wu:2012hs}.

Since in the present work we are interested in the gravitational waves, which are detected
through the change of line element, we only need to focus on the components of tetrad
corresponding to the components of metric. In particular, decomposing the tetrad as
\cite{Wu:2012hs}
\begin{equation}
 e^A_{\mu}(x) = \bar{e}^A_{\mu}(x) + \chi^A_{\mu}(x) ~,
\end{equation}
which satisfies the condition
\begin{equation}
\label{decompose}
 g_{\mu\nu}(x) = \eta_{AB} e^A_{\mu} e^B_{\nu} = \eta_{AB} \bar{e}^A_{\mu} \bar{e}^B_{\nu}
~,
\end{equation}
where $\bar{e}^A_{\mu}$ illustrates the part of tetrad corresponding to metric components
and $\chi^A_{\mu}$ represents the degrees of freedom released from the local Lorentz
transformation (whose number is thus six), we only need to focus on the $\bar{e}^A_{\mu}$
part.

As usual, we perturb the tetrad fields $\bar{e}^A_{\mu}$ around a flat
Friedmann-Robertson-Walker (FRW) background as follows,
\begin{align}
\label{perturbation of tetrads}
 \bar{e}^0_{\mu} =& \delta^0_{\mu} (1+\psi) + a \delta^i_{\mu} (G_i+\partial_i F) ~,
\nonumber \\
 \bar{e}^a_{\mu} =& a \Big[ \delta^a_{\mu}(1-\phi) \nonumber \\
  	& \ \ \ \, + \delta^i_{\mu} \delta^{aj} \Big( \frac{1}{2} h_{ij} + \partial_i
\partial_j B + \partial_j C_i + \partial_i C_j \Big) \Big] ~, \nonumber \\
 \bar{e}_0^{\mu} =& \delta_0^{\mu} (1-\psi) - \frac{1}{a} \delta^{\mu i} (G_i+\partial_i
F) ~, \nonumber \\
 \bar{e}_a^{\mu} =& \frac{1}{a} \Big[ \delta_a^{\mu} (1+\phi) \nonumber \\
  	& \ \ \ \, - \delta^{\mu i} \delta^j_a \Big( \frac{1}{2} h_{ij} + \partial_i
\partial_j B + \partial_i C_j + \partial_j C_i \Big) \Big] ~,
\end{align}
where $a(t)$ is the scale factor, and with small Latin indices from the beginning of the
alphabet spanning the spatial part of the tangent space. In the above expressions we have
introduced the scalar modes $\phi$ and $\psi$, the transverse vector modes $C_{i}$ and
$G_{i}$, and the transverse traceless tensor mode $h_{ij}$.

The above perturbed tetrad gives rise to the standard perturbed FRW metric
\begin{align}
 g_{00} = & -1-2\psi ~, \nonumber\\
 g_{i0} = & -a [\partial_i F+G_i] ~,  \\
 g_{ij} = & a^2 [ (1-2\phi) \delta_{ij} + h_{ij} + \partial_{i} \partial_j B + \partial_j
C_i + \partial_i C_j ] ~.\nonumber
\end{align}
We mention that the above perturbation expansions are slightly different from those
provided in \cite{Chen:2010va}, since there are more than one forms of tetrads
that correspond to the same metric.

In the rest of the manuscript we focus on the GWs, i.e. the tensor perturbations.
Correspondingly, from now on, we set the scalar and vector perturbations to zero for
convenience. Inserting (\ref{perturbation of tetrads}) into (\ref{torsiontensor}) we
obtain the perturbed torsion tensor as
\begin{align}
 T^i{}_{0j} = & H\delta_{ij} + \frac12 \dot{h}_{ij} \nonumber \\
 T^i{}_{jk} = & \frac12 \left( \partial_j h_{ik} - \partial_k h_{ij} \right) ~,
\end{align}
where $H\equiv\dot{a}/a$ is the Hubble function. As a result, the torsion scalar from
(\ref{torsiscal}) reads as \footnote{Mind the sign difference comparing to the majority
of $f(T)$ works, due to the mostly-plus signature we use in this manuscript.}
\begin{equation}
T =T^{(0)}+O(h^2)= 6H^2 +O(h^2) ~,
\label{Texpanss}
\end{equation}
where $T^{(0)}$ is the zeroth-order part of the torsion scalar expansion (from now on the
superscript ${}^{(0)}$ marks the   zeroth-order part of an expanded quantity).
Thus, relation (\ref{Texpanss}) implies that the torsion scalar is not affected by the
tensor fluctuations at linear expansion. Additionally, from (\ref{superpot}) we acquire
the perturbed
super-potential as
\begin{align}
 S_i{}^{0j} &= H\delta_{ij} -\frac14\dot{h}_{ij} \notag \\
 S_i{}^{jk} &= \frac{1}{4a^2} (\partial_jh_{ik} -\partial_k h_{ij}) ~.
\end{align}

Inserting the above perturbations in the field equations (\ref{fieldeqs}) we can obtain
\begin{align}
 & 4 f_T \Big[ \big( \dot{H}+3H^2 \big) \delta_{ij}+\frac14 \big(
-\ddot{h}_{ij}+\frac{\nabla^2}{
a^2}{h}_{ij}-3H\dot{h}_{ij} \big) \Big] \nonumber \\
 & + 4\dot{f}_T \big( H\delta_{ij}-\frac{\dot{h}_{ij}}{4} \big) -f\delta_{ij}=16\pi G
\Theta^{i}{}_{
j} ~,
\end{align}
where the derivative $f_{T}$ is calculated at
$T=T^{(0)}=6H^2$. The perturbation part of the above equation leads to the equation of
motion
for the
GWs, namely
\begin{equation}
\label{eom}
 \ddot{h}_{ij} + \big( 3H +\frac{\dot{f}_T}{f_T} \big) \dot{h}_{ij}
-\frac{\nabla^2}{a^2}{h}_{ij}=0
~.
\end{equation}

\section{Gravitational waves in $f(T)$ gravity via the EFT approach}
\label{EFTanalysis}

In this section we will investigate the GWs in $f(T)$ gravity through the EFT approach
\cite{ArkaniHamed:2003uy}. Inspired by the theory of spontaneous symmetry breaking of the
$SU(2)\times U(1)$ gauge theory of the Standard model, one can apply the EFT to
cosmological perturbations of modified gravity theories, by treating them as the Goldstone
boson of spontaneously broken time-translations. Similarly to  the gauge field theory with
spontaneous symmetry breaking, one can also choose the unitary gauge, ``eating'' the
would-be Goldstone bosons and making the theory to display only metric degrees of
freedom.
A significant advantage is that this process can organize the terms of the action as
number of perturbations, allowing us to deal with the background and perturbations
separately \cite{ArkaniHamed:2003uy}.
The EFT approach has been applied to the inflationary context \cite{Creminelli:2006xe,
ArkaniHamed:2007ky, Cheung:2007st} or to the dark energy paradigm
\cite{Creminelli:2008wc, Gubitosi:2012hu,
Bloomfield:2012ff, Bloomfield:2013efa, Li:2018} (see also \cite{Creminelli:2017sry} for an
EFT analysis of dark energy models in the light of GW170817).

\subsection{The description of EFT approach}

Let us first describe the EFT approach. For simplicity we use the curvature-based
formulation of
gravity, and we start by considering a general action of the form \cite{Gubitosi:2012hu}
{\small{
\begin{align}
\label{curvature action}
 S =&\int d^4x \Big\{ \sqrt{-g} \Big[ \frac{M^2_P}{2} \Omega(t)R - \Lambda(t) -
b(t)g^{00} \nonumber  \\
 & \
 + M_2^4 (\delta g^{00})^2 - \bar{m}^3_1 \delta g^{00} \delta K - \bar{M}^2_2 \delta K^2
-\bar{M}^2_
3\delta K^{\nu}_{\mu}\delta K^{\mu}_{\nu} \nonumber \\
 & \
 + m^2_2 h^{\mu\nu} \partial_{\mu} g^{00} \partial_{\nu} g^{00}\! +\! \lambda_1 \delta
R^2\! + \!\lambda_2 \delta R_{\mu\nu} \delta R^{\mu\nu} \!+\! \mu^2_1 \delta g^{00}
\delta R \Big] \nonumber \\
 & + \gamma_1 C^{\mu\nu\rho\sigma} C_{\mu\nu\rho\sigma} + \gamma_2
\epsilon^{\mu\nu\rho\sigma} C_{\mu\nu}^{\quad\kappa\lambda} C_{\rho\sigma\kappa\lambda}
\nonumber \\
 & + \sqrt{-g} \Big[ \frac{M^4_3}{3} (\delta g^{00})^3 - \bar{m}^3_2 (\delta g^{00})^2
\delta K + ..
. \Big] \Big\} ~,
\end{align}}}
where $C^{\mu\nu\rho\sigma}$ is the Weyl tensor, $\delta K^{\nu}_{\mu}$ is the
perturbation of the extrinsic curvature and $R$ is the Ricci scalar corresponding to the
Levi-Civita connection. Additionally, we have included the time-dependent functions
$\Omega(t)$, $\Lambda(t)$, $b(t)$ which are determined by the background evolution, and
finally we have allowed for various time-dependent coefficients in front of the
various terms. We mention here that the first line of the action corresponds to the
background evolution, the lines from second to fourth are quadratic in
perturbations, while the fifth line is cubic in perturbations.

\subsection{The EFT of teleparallel gravity}

Now we proceed to the application of the EFT approach in TEGR and the
modified theory based on this framework \cite{Li:2018}, an application which is
facilitated by the fact that teleparallel gravity can be seen as a translational gauge
theory of gravity \cite{Pereira:book}. As we will see, in order to do this we need to add
some extra terms to the above action, both at the background and perturbation parts.

We begin by referring to the unitary gauge. In a general perturbed FRW geometry, a scalar
degree of freedom is decomposed as
\begin{equation}
 \phi(t,\vec{x}) = \phi_0(t) + \delta\phi(t,\vec{x}) ~.
\end{equation}
The unitary gauge is to choose the coordinate $t$ to be a function of $\phi$, namely
$t=t(\phi)$,
thus $\delta\phi=0$ and the action displays only metric degrees of freedom.

The unitary gauge action must be invariant under the unbroken symmetries. This implies
that the action should leave spatial diffeomorphisms unbroken. Thus, having relations
(\ref{LR}),(\ref{KK}) and (\ref{RT}) in mind, it is reasonable to include both curvature
and torsion terms in the action. In summary, the action of EFT can contain
\cite{Piazza:2013coa}:
\begin{itemize}
\item i) Terms that are invariant under all diffeomorphisms. These are four-dimensional
diffeomorphisms invariant scalars such as $R$ and $T$, which are in general multiplied by
functions of time.

\item ii) Terms that are  invariant only under spatial diffeomorphisms. Firstly, these
can be scalars that are constructed by spatial tensors such as the spatial Riemann tensor
$^{(3)}R_{\mu\nu\rho\sigma}$, the extrinsic curvature $K_{\mu\nu}$, as well as the the
spatial torsion tensor $^{(3)}
T^{\rho}_{\ \mu\nu}$ and the ``extrinsic torsion'' $\hat{K}_{\mu\nu}$. We will explain
the latter two terms more clearly below. Secondly, these can be four-dimensional
covariant tensors with upper $0$ indices such as $g^{00}$, $R^{00}$ and $T^{0}$
($T^{0}$ is the 0-index component of the contracted torsion tensor $T^{\mu}$).
\end{itemize}

The terms of type ii) arise from the definition of a preferred time slicing by the scalar
field $\phi$, namely
\begin{equation}	
 n_{\mu} = \frac{ \partial_{\mu} \phi(t) }{ \sqrt{-(\partial{\phi})^2}} =
\frac{\delta^0_{\mu}}{\sqrt{-g^{00}}} ~.
\end{equation}
Since the time-translation is broken, we can contract covariant tensors with this unitary
vector orthogonal to the $t = const$ surfaces, and this is where the terms with
upper $0$ indices arise from. Then we consider the Weitzenb\"ock covariant derivative of
$n_{\mu}$, projected on the surface of constant $t$, i.e.
\begin{equation}
 h^{\sigma}_{\mu} \hat{\nabla}_{\sigma} n_{\nu} \equiv \hat{K}_{\mu\nu} ~,
\end{equation}
where $h_{\mu\nu} \equiv g_{\mu\nu} +n_{\mu\nu}$ is the induced metric of this surface.
It is easy to verify that this quantity is a spatial tensor, therefore we refer to it as
``extrinsic torsion''. Given the relation (\ref{DR}) between the Weitzenb\"ock covariant
derivative and the ordinary covariant derivative, we can extract the following relations
between extrinsic curvature and extrinsic torsion:
\begin{equation}
\label{Kmunu}
 \hat{K}_{\mu\nu} \equiv h^{\sigma}_{\mu} \hat{\nabla}_{\sigma} n_{\nu} = K_{\mu\nu} -
K^{\lambda}_{
\ \nu\mu} n_{\lambda} + n_{\mu} \frac{1}{g^{00}} T^{00}_{\quad\nu} ~.
\end{equation}
After contracting its two indices, we have
\begin{align}
\label{K}
 K =& \nabla_{\mu}n^{\mu} = \hat{\nabla}_{\mu}n^{\mu}-K^{\mu}_{\ \lambda\mu}n^{\lambda}
\nonumber \\
 =& \hat{K} +T_{\lambda} n^{\lambda} = \hat{K} + (-g^{00})^{-1/2} T^0 ~.
\end{align}
Accordingly, having in mind (\ref{Kmunu}) and (\ref{K}) and observing the action
(\ref{curvature action}), we
can deduce that, if we allow $T^{00\nu}$ and $T^0$ to be present in our action, we can
avoid the
use of $\hat{K}$ and $\hat{K}_{\mu\nu}$.

We proceed by considering the covariant derivative of $n_{\mu}$ perpendicular to the time
slicing, which gives
\begin{equation}	
 n^{\sigma} \hat{\nabla}_{\sigma} n_{\nu} = n^{\sigma} \nabla_{\sigma} n_{\nu} +
\frac{1}{g^{00}}T^{
00}_{\quad \nu} ~.
\end{equation}
As illustrated in \cite{Cheung:2007st}, the first term on the right-hand side just leads
to a term that contains $g^{00}$ and $h^{\mu}_{\nu}$, which has already been included in
action (\ref{curvature action}). Therefore, we can also avoid $n^{\sigma}
\hat{\nabla}_{\sigma} n_{\nu}$ if we allow for $T^{00\nu}$ in the action.

Using the unit normal vector $n_{\mu}$ (or the projection operator $h^{\mu}_{\nu}$), we
can construct three-dimensional spatial tensors, whose contractions provide the spatial
diffeomorphisms invariant scalar, which can then be used in the action. On the other
hand, for convenience we only use four-dimensional tensors, since their three-dimensional
counterparts can always be expressed by using the normal vector or the projection
operator, for instance as
\begin{align}
 ^{(3)}R_{\mu\nu\rho\sigma} = h^{\alpha}_{\mu} h^{\beta}_{\nu} h^{\gamma}_{\rho}
h^{\delta}_{\sigma}
 R_{\alpha\beta\gamma\delta} - K_{\mu\rho} K_{\nu\sigma} + K_{\nu\rho} K_{\mu\sigma} ~.
\end{align}

Finally, since the spatial covariant derivative of a spatial tensor can be obtained as
the projection of the four-dimensional covariant derivative, it is implied that the use
of spatial covariant derivatives can also be avoided.

We proceed by arranging all the above kinds of operators in powers of number of
perturbations. Hence, we focus on the expansion around the FRW background. If we consider
an operator composed by the contraction of two tensors $X$ and $Y$ (it is straightforward
to generalize to more tensors), expanded linearly as  $X\approx X^{(0)}+\delta X$ and
$Y\approx Y^{(0)}+\delta Y$,  then we can expand it as
\begin{align}
\label{TGexpr}
 XY = \delta X \delta Y + X^{(0)}Y + XY^{(0)} - X^{(0)} Y^{(0)} ~.
\end{align}
Given the FRW background, the unperturbed tensors $X^{(0)}$ and $Y^{(0)}$ can then always
be expressed as functions of $g_{\mu\nu}$, $n_{\mu}$ and $t$. Hence, in the case of
Riemann and torsion tensors, and of extrinsic curvature and torsion, we can write
\begin{align}
\label{Rexpr1}
 R_{\mu\nu\rho\sigma}^{(0)} &= f_{1}(t) g_{\mu\rho} g_{\nu\sigma} + f_2(t) g_{\mu\rho}
n_{\nu} n_{\sigma} + f_3(t) g_{\mu\sigma} g_{\nu\rho} \nonumber \\
 & \ \ \ +f_4(t) g_{\mu\sigma} n_{\nu} n_{\rho} + f_5(t) g_{\nu\sigma} n_{\mu} n_{\rho}
\nonumber \\
 &\ \ \ +f_6(t) g_{\nu\rho} n_{\mu} n_{\sigma}, \\
\label{Rexpr2}
 T^{(0)}_{\rho\mu\nu} &= g_1(t) g_{\rho\nu} n_{\mu} + g_2(t) g_{\rho\mu} n_{\nu}, \\
\label{Rexpr3}
 K^{(0)}_{\mu\nu} &= f_7(t) g_{\mu\nu} + f_8(t) n_{\mu} n_{\nu}, \\
\label{Rexpr4}
 \hat{K}^{(0)}_{\mu\nu} &=0 ~.
\end{align}
Since the last term $X^{(0)}Y^{(0)}$ in (\ref{TGexpr}) is merely a polynomial of $g^{00}$
with time dependent coefficients, we mention that expressions
(\ref{Rexpr1})-(\ref{Rexpr4}) hold modulo a factor of polynomials of $g^{00}$, which is
irrelevant for our analysis.

Now, the first term in (\ref{TGexpr}) starts explicitly quadratic in perturbations and
hence we
keep it. Concerning the second term, namely $X^{(0)}Y$, by construction $Y$ will be
linear in $K_{\mu\nu}$, $R_{\mu\nu\rho\sigma}$, $\hat{K}_{\mu\nu}$ and $T^{\rho}_{\
\mu\nu}$, with covariant
derivatives acting on them. Due to the relation (\ref{LR}) between the two kinds of
connection, we can consider only the covariant derivative with respect to the Levi-Civta
connection, which can be dealt with by successive integration by parts, allowing them
tact
on $X^{(0)}$ and the time dependent coefficient. This process will generate extrinsic
curvature terms. Hence, after contracting all the indices, we will obtain the only
possible scalar linear terms with no covariant derivatives: $K$, $R^{00}$, $R$,
$\hat{K}$, $T^0$, $T$. Given the relation (\ref{K}), $\hat{K}$ can be
eliminated in terms of $T^0$. As demonstrated in \cite{Cheung:2007st}, the integration of
$K$ and
$R^{00}$ with time dependent coefficients  results to just the linear operator $g^{00}$
plus
invariant terms that start quadratically in the perturbations, which implies that these
two terms
can also be avoided in the background action.

Finally, by observing relation (\ref{RT}), one deduces that the integration of the
boundary term with a time-dependence coefficient takes the form of
\begin{align}
 S = \int d^4x \sqrt{-g} f(t) \nabla_{\mu} T^{\mu}
 = -\int d^4x \sqrt{-g} \dot{f}(t)T^0 ~.
\end{align}

Having all the above discussions in mind, we can now write down the EFT action of
teleparallel gravity. As we showed, the remaining background terms
are $R$ and $T^0$, and hence we have
\begin{align}
 S = & \int d^4x \sqrt{-g} \Big[ \frac{M^2_P}{2} \Omega(t)R - \Lambda(t) - b(t) g^{00} +
\frac{M^2_
P}{2} d(t) T^0 \Big] \nonumber \\
 & \ \ \ \ + S^{(2)} ~,
 \label{actionfin}
\end{align}
with $d(t)$ a time-dependent function. In the second line of the above action, $S^{(2)}$
has been introduced to include all the terms that start explicitly quadratic in the
perturbations, and hence its presence does not affect the background dynamics. In addition
to the terms shown in action (\ref{curvature action}), $S^{(2)}$ can also include:
i) Pure torsion terms such as $\delta T^2$, $\delta T^0\delta T^0$ and $\delta
T^{\rho\mu\nu}\delta
T_{\rho\mu\nu}$ (note that since we include $T^{0}$ in the action, according to (\ref{K})
we can avoid the presence of $\hat{K}$);
ii) Terms mixing torsion and curvature such as $\delta g^{00}\delta T$, $\delta
g^{00}\delta T^0$,
$\delta T\delta R$ and $\delta K\delta T^0$.

\subsection{The EFT of $f(T)$ gravity}

In the previous subsection we applied the EFT approach to teleparallel equivalent of
general relativity. Thus, we have all the machinery to proceed to the
EFT approach of $f(T)$ gravity.

A first complication that arises from such project is the incorporation of the time
slicing. Similarly to the discussion of $f(R)$ theory within EFT formalism
\cite{Gubitosi:2012hu}, we firstly expand the action (\ref{fT}) with respect to the
background as
\begin{align}
 S =& \frac{M^2_P}{2} \int d^4x \sqrt{-g} \Big[ f_{T}T +f(T^{(0)}) -f_{T}T^{(0)}
\nonumber \\
 & ~~~~ ~~~~ \ \ \ \ \ \ \ \ \ \ \, \ \ \  +\frac12 f_{TT}\delta T^2+... \Big] ~.
\end{align}
Afterwards, we can fix the time slicing in a way that it coincides with uniform $T$
hypersurfaces. This treatment will make the terms beyond the linear order in the above
expansion to vanish, since their contribution to the equations of motion will always
include at least one power of $\delta T$. Thus, we obtain the unitary-gauge action as
follows
\begin{equation}
\label{action}
 S = \frac{M^2_{P}}{2} \int d^4x \sqrt{-g} \big[ -f_TR -2\dot{f_T} T^0 -T^{(0)} f_T
+f(T^{(0)}) \big] ,
\end{equation}
which comparing with (\ref{actionfin}) then implies:
\begin{align}
 \Omega(t) &= -f_T(T^{(0)}) ~,~~
 d(t) = -2\dot{f}_T(T^{(0)}) ~,\nonumber \\
 \Lambda(t) &= \frac{M^2_{*}}{2}\left[T^{(0)}f_T(T^{(0)})-f(T^{(0)})\right] ~,~~
 b(t) = 0 ~.
\end{align}
Note that from (\ref{action}) and (\ref{actionfin})  we also deduce that $S^{(2)}=0$,
and thus one only needs to deal with the background part.
 Lastly, if the additional $T^0$ term vanishes, then the above action will reproduce the
EFT form
of $f(R)$ gravity \cite{Gubitosi:2012hu}.

\subsection{Propagations of GWs in $f(T)$ gravity from EFT}

We have now all the tools in order to proceed to the investigation of GWs in the EFT
approach. As we mentioned earlier, in order to study the GWs we only need to focus on the
$h_{ij}$ component of the tetrad. Additionally, we mention that the quantities in
action (\ref{action}) must be expanded to quadratic order in perturbations in order to
obtain the dynamical behavior. Thus, the perturbative components of the metric, up to
second order in perturbations, read
\begin{align}
\label{DP}
 &g_{00} = -1 ~,~~ g_{0i} = 0~, \nonumber\\
 &g_{ij} = a^2 \big( \delta_{ij}+h_{ij}+\frac{1}{2}h_{ik}h_{kj} \big) ~,
\end{align}
which can be derived from the perturbative tetrads (up to second order in perturbations as
well):
\begin{align}
 &\bar{e}^0_{\mu} = \delta^0_{\mu} ~, \\  	
 &\bar{e}^a_{\mu} = a\delta^a_{\mu} +\frac{a}{2} \delta^i_{\mu} \delta^{aj} h_{ij} +
\frac{a}{8}
 \delta^i_{\mu} \delta^{ja} h_{ik} h_{kj} ~, \\
 &\bar{e}_0^{\mu} = \delta_0^{\mu} ~, \\
 &\bar{e}_a^{\mu} = \frac{1}{a} \delta^{\mu}_a -\frac{1}{2a} \delta^{\mu i} \delta^j_a
h_{ij} + \frac{1}{8a} \delta^{i\mu} \delta^j_a h_{ik} h_{kj} ~.
\end{align}
Accordingly, we can calculate $T^0$ and find that its perturbation part vanishes up to
second order, which is given by
\begin{align}
 T^0 = -T_0 =e^{\lambda}_A\partial_0
e^A_{\lambda}-e^{\lambda}_A\partial_{\lambda}e^A_0 = 3H ~.
\end{align}
As a result, the $T^0$ term in   action (\ref{action}) does not lead to a new kinetic
term. This feature lies beyond the main result of the present work.

With the metric provided in (\ref{DP}) we calculate $R$ as follows
\begin{equation}
\label{R}
 R = {}^{(3)}R + K_{\mu\nu} K^{\mu\nu} -K^2 + 2 \nabla_{\mu} (Kn^{\mu} -n^{\rho}
\nabla_{\rho} n^{\mu} ) ~,
\end{equation}
and up to second order we have the following expressions:
\begin{align}
 ^{(3)}R &\approx -\frac{1}{4} a^{-2} \big( \partial_i h_{kl} \partial_i h_{kl} \big) ~,
\\
 K^{ij} K_{ij} &\approx 3H^2 +\frac{1}{4} \dot{h}_{ij} \dot{h}_{ij} ~, \\
 K &\approx 3H ~,
\end{align}
with $^{(3)}R$ being the spatial curvature scalar. We mention that when a scalar field
$\phi$ is
non-minimally coupled to $R$, the total derivative in (\ref{R}) does not vanish. Thus, we
should
consider its integral with a time-dependent coefficient, which is given by
\begin{align}
 \int & d^4x \sqrt{-g} f(t) 2\nabla_{\nu} \big( Kn^{\mu} -n^{\mu} \nabla_{\mu} n^{\nu}
\big) \\
 & = \int d^4x \sqrt{-g} \big( -2 \dot{f} K n^0 \big) = \int d^4x \sqrt{-g} \big( 6H
\dot{f} \big)
~. \nonumber
\end{align}
Hence, we deduce that this term does not contribute to tensor perturbations up to second
order.

As a result, we obtain the final form of the action (\ref{action}) for the linearized GWs
within a
cosmological background as follows:
\begin{align}
 \!\!\!\!S = & \frac{M^2_{P}}{2} \int d^4x \sqrt{-g} \Big[ \frac{f_T}{4}
\big(a^{-2} \vec{\nabla} h_{ij} \cdot \vec{\nabla} h_{ij} - \dot{h}_{ij} \dot{h}_{ij}
\big) \nonumber \\
 & \ \ \ \ \ \ \ \  + 6H^2 f_T - 12 H \dot{f}_T -T^{(0)} f_T +f(T^{(0)}) \Big] ~.
\end{align}
The above action is exactly the EFT action of cosmological GWs within the $f(T)$ gravity
up to second order. Varying this action with respect to $h_{ij}$ can again yield
the equation of motion (\ref{eom}). Then, performing the Fourier
transformation and  tracing the evolution of a fixed Fourier mode of GWs we obtain
\begin{equation}
\label{GWeq2momspace}
 \ddot{h}_{ij} +3H \big( 1-\beta_T \big) \dot{h}_{ij}+\frac{k^2}{a^2} h_{ij} =0 ~,
\end{equation}
where we have introduced the dimensionless parameter
\begin{eqnarray}
 \beta_T \equiv -\frac{\dot{f}_T}{3Hf_T} ~.
 \label{betadef}
\end{eqnarray}
 Observing Eq. (\ref{GWeq2momspace}), and  comparing it with the general evolution
equation  of linear, transverse-traceless perturbations over an FRW background,
 we can immediately deduce that the speed of GWs
is equal to one, i.e. equal to the speed of light.
As a result, one can see that the experimental constraint of GW170817 on the GW speed
in $f(T)$ gravity is trivially satisfied.

We can proceed with our analysis by referring to the dispersion relation
and the frequency of cosmological GWs in $f(T)$ gravity. Taking the ansatz of the Fourier
transformation of cosmological GWs as
\begin{equation}
 h_{ij} = \int d^3 k e^{i\vec{k}\cdot\vec{x}} \big[ A_{ij}e^{i\omega t } + B_{ij}
e^{-i\omega t} \big] ~,
\end{equation}
and inserting it into (\ref{eom}), we obtain
\begin{equation}
 \Big( \frac{k^2}{a^2} - \omega^2 \Big) \pm 3i{H} \omega \Big( 1 +\frac{\dot{f}_T}{3H f_T}
\Big) =
0 ~.
\end{equation}
The solution of the above equation leads to the dispersion relation, which can be
expressed as
\begin{equation}
 \left| {\frac{d\omega}{dk}} \right|= \frac{1}{a} \Big[ 1 -\frac{9a^2}{4k^2} H^2
(1-\beta_T)^2 \Big]^{
-\frac{1}{2}} ~.
\end{equation}
Note that the dimensionless coefficient $\beta_T$ from (\ref{betadef}), using that
$T=6H^2$ can be
further written as
\begin{align}
 \beta_T = \frac{d \ln f_T}{d\ln T}(1+w_{tot}) ~,
\end{align}
with $w_{tot}\equiv  -1 -\frac{2\dot{H}}{3H^2}$
the total equation-of-state parameter of the universe ($w_{tot} = p_{tot}/\rho_{tot}$ with
$p_{tot}$
and $\rho_{tot}$
the total pressure and energy density of the universe respectively).

It is straightforward to notice that if the gravitational theory is GR or TEGR, then $\beta_T=0$. However, for a general case of $f(T)$ gravity, the form of $\beta_T$ deviates from zero. Therefore, even if the propagation of GWs in $f(T)$ gravity remains at the speed of light, similarly to general relativity, a high precision measurement of the dispersion relation of cosmological GWs can impose observational constraints upon the parameter $\beta_T$, and hence reveal the effect of $f(T)$ gravity. Definitely, since in viable $f(T)$ models $f_{TT}\ll1$ \cite{Iorio:2012cm, Nesseris:2013jea, Nunes:2016qyp, Wu:2018}\footnote{Using the best-fit values for the parameters of specific $f(T)$ models \cite{Nesseris:2013jea, Nunes:2016qyp} we obtain $\beta_T\sim10^{-2}-10^{-1}$ in the low-redshift region.}, we deduce that it is quite difficult to utilize the present GWs experiments to probe such a deviation in the dispersion relation. However, a future measurement of the value of the $\beta_T$ parameter may become possible under the great development of GW astronomy \cite{Zhao:2017imr, Caprini:2018mtu} (note that $\beta_T$ is related to the running of the effective Planck mass \cite{Ezquiaga:2017ekz, Baker:2017hug} but it does not coincide with it). If a non-zero $\beta_T$ is measured in future observations, it could be the smoking gun of modified gravity, and the significance of this signature would become even greater if one has in mind that probably there is no impact of $f(T)$ gravity upon the polarization modes of GWs when comparing to the case of GR \cite{Bamba:2013ooa, Abedi:2017jqx}.

\section{Conclusions}
\label{Seciontconclusions}

In this work we performed a detailed analysis of the GWs in $f(T)$ gravity and cosmology. Taking advantage of the fact that teleparallel gravity can be seen as a translational gauge theory of gravity, we applied the EFT approach, which allows to analyze the perturbations in a systematic way and separately from the background evolution. Constructing all terms in the perturbative action up to second order, we extracted the propagation equation for the GWs. For completeness, we alternatively extracted the same equation through the standard scalar vector and tensor perturbation analysis around a cosmological background.

From the GW propagation equation we deduced that the speed of GWs in $f(T)$ gravity is equal to the light speed. This is the main result of the present work and it is very important since it shows that $f(T)$ gravity can trivially satisfy the combined constraints of GW170817 and GRB170817A. Note that this is not guaranteed in a general theory of modified gravity, in which GWs propagate with a speed that may be different from the speed of light. Therefore, it is necessary to examine the observational constraints upon the propagation of cosmological GWs in modified gravities. The above result offers an additional advantage for $f(T)$ gravity.

Finally, examining the dispersion relation and the frequency of the GWs in $f(T)$ gravity within an FRW background, we found that there is a deviation from the result of GR, which is quantified by a new parameter $\beta_T$, due to a modification of the friction term in the perturbation equation of cosmological GWs. Although for $f(T)$ models that are allowed by present observations the value of $\beta_T$ is typically small and is difficult to be tested in the current GW data, a possible future measurement in advancing GW astronomy would be the smoking gun of testing this type of modified gravity.

\begin{acknowledgments}
We are grateful to Yong Cai, Ignacy Sawicki, Yang Zhang and Wen Zhao for helpful discussions.
YFC, CL and LQX are supported in part by the Chinese National Youth Thousand Talents Program, by the NSFC (Nos. 11722327, 11653002, 11421303, J1310021), by the CAST Young Elite Scientists Sponsorship Program (2016QNRC001), and by the Fundamental Research Funds for the Central Universities. 
YFC and LQX are grateful to ENS for his hospitality at the National Technical University of Athens, Greece during the preparation of this work.
This article is based upon work from COST Action ``Cosmology and Astrophysics Network for Theoretical Advances and Training Actions'', supported by COST (European Cooperation in Science and Technology). 
Part of numerical simulations are operated on the computer cluster LINDA in the particle cosmology group at USTC.
\end{acknowledgments}

\end{document}